\documentclass[10pt,sigconf,letterpaper]{acmart-PL}

\usepackage{soul}
\usepackage{todonotes}
\usepackage{url}
\usepackage{graphicx}
\graphicspath{ {img/} }

\usepackage{ulem}
\usepackage{xcolor}
\usepackage{subfigure}
\usepackage{enumitem}
\usepackage{diagbox}
\usepackage{colortbl}
\usepackage{array}
\newcolumntype{P}[1]{>{\centering\arraybackslash}p{#1}}

\usepackage{tikz}

\newcommand{\ie}{{\textit i.e.,\/ }}
\newcommand{\eg}{{\textit e.g.,\/ }}

\begin{document}

\title{GossipSub: Attack-Resilient Message Propagation in the Filecoin and ETH2.0 Networks}



\author{Dimitris Vyzovitis}
 \affiliation{Protocol Labs}
\email{vyzo@protocol.ai}

\author{Yusef Napora}
 \affiliation{Protocol Labs}
\email{yusef@protocol.ai}

\author{Dirk McCormick}
 \affiliation{Protocol Labs}
\email{dirk@protocol.ai}

\author{David Dias}
 \affiliation{Protocol Labs}
\email{david@protocol.ai}



\author{Yiannis Psaras}
\affiliation{Protocol Labs}
\email{yiannis@protocol.ai}






\begin{abstract}

Permissionless blockchain environments necessitate the use of a fast and attack-resilient message propagation protocol for Block and Transaction messages to keep nodes synchronised and avoid forks. We present GossipSub, a gossip-based pubsub protocol, which, in contrast to past pubsub protocols, incorporates resilience against a wide spectrum of attacks.
Firstly, GossipSub's \textit{mesh construction implements an eager push model} keeps the fan-out of the pubsub delivery low and balances excessive bandwidth consumption and fast message propagation throughout the mesh. Secondly, through \textit{gossip dissemination, GossipSub realises a lazy-pull model} to reach nodes far-away or outside the mesh. Thirdly, through constant observation, \textit{nodes maintain a score profile for the peers they are connected to}, allowing them to choose the most well-behaved nodes to include in the mesh. Finally, and most importantly, a number of tailor-made mitigation strategies designed specifically for these three components make GossipSub resilient against the most challenging Sybil-based attacks. We test GossipSub in a testbed environment involving more than 5000 VM nodes deployed on AWS and show that it stays immune to all considered attacks. GossipSub is currently being integrated as the main messaging layer protocol in the Filecoin and the Ethereum 2.0 (ETH2.0) blockchains.


\end{abstract}

\maketitle

\section{Introduction}

Blockchain environments require a message layer protocol to propagate transaction and block messages. Transaction messages are those that transfer  value, normally in the form of tokens or crypto-assets between different nodes in the network.
Block messages are produced by a subset of nodes, the ``miner'' nodes, to group sets of transactions together, update the state of the blockchain and keep all nodes in sync.
Being up-to-date, or synchronised, with regard to both transaction and block messages is a very important requirement in blockchain networks and it is the task of the message propagation protocol to achieve. In the opposite case, nodes have differing views of the state of the network, are working on different data-sets and cause the blockchain to fork.

Furthermore, open blockchain networks, such as Bitcoin \cite{6688704}, Ethereum \cite{8802401} and Filecoin \cite{filecoin} are \textit{``permissionless''} with no central control by any entity or organisation to oversee and manage access control (\ie the identity of nodes joining the network). The network is an open, unstructured peer-to-peer (P2P) network and can be joined by any node that claims to run the protocol. Malicious nodes can, therefore, join and try to distort the view of nodes with regards to latest transactions and blocks with the aim of disrupting the network or stealing monetary value. Malicious activity is most commonly realised through Sybil nodes \cite{10.1007/3-540-45748-8_24} that can carry out a range of attacks, the most prominent of which is the eclipse attack \cite{10.1145/1133572.1133613, 190890} \ie silencing of nodes in order to cause packet loss and in turn, blockchain forks. Given that those networks carry large amounts of monetary value, potentially in the order of hundreds of millions of dollars, they are very attractive to attackers. Therefore, \textit{the messaging layer protocol of a blockchain network has to be resilient against malicious nodes that attack the network by dropping or delaying messages.}


The most straightforward strategy to alleviate both message propagation speed and resilience/security concerns in permission-less, unstructured P2P networks is \textit{flooding} - used in the Bitcoin network \cite{6688704, 10.1145/3319535.3354237}. Flooding ensures that messages propagate as fast as possible throughout the network and is robust against eclipse attacks, as it introduces high levels of traffic redundancy.
Flooding, however, is very expensive in terms of bandwidth.
Authors in \cite{10.1145/3319535.3354237} report that up to 44\% of the overall traffic of the Bitcoin network is redundant. This figure translates to a bandwidth requirement of up to 350GBs/month for a public Bitcoin node and for the current network size.

``Gossip''-based pubsub protocols have been introduced in the past as a way to limit the number of messages propagated between peers in pub/sub systems \cite{Leitao:2007:HMP:1251984.1253089}, \cite{Baldoni:2007:TTE:1266894.1266898}, \cite{4365705}. In gossip-based approaches, peers forward metadata related to messages they have ``seen'' without forwarding the messages themselves, a method generally called \textit{lazy pull}. However, given that past systems were either centralised from the point of view of (ownership and) access control, or were not carrying any commercial/financial weight, \textit{they were not designed to be resilient against attacks.}

In this paper, we are proposing GossipSub, a gossip-based pubsub protocol that is designed to deal with both fast and resilient message propagation in permissionless networks. GossipSub consists of two main components, the \textit{Mesh Construction} and the \textit{Score Function}, as well as a set of mitigation mechanisms designed exclusively on top of the two aforementioned protocol components. As we show here, the combination of these parts render the protocol resilient against a wide-range of attacks, responsive to dynamic network conditions and fast in terms of message propagation.
\begin{enumerate}[leftmargin=*]
    \item \textbf{The Mesh Construction:} GossipSub introduces a connected \textit{global mesh} structure, where each node is connected to a limited number of other peers forming its \textit{local (view of the) mesh}. Mesh-connected nodes directly share messages with one another, realising an \textit{eager push} communication model. Nodes can join and leave the mesh based on network-level conditions or application-level semantics. Those nodes that are not part of the mesh communicate with mesh-connected nodes through gossip (\ie \textit{lazy push}).
    \item \textbf{The Score Function:} Every node participating in the network is \textit{constantly being observed by all of the nodes it is connected to} (whether in the same local mesh or not). We have carefully selected a number of actions that either flag malicious activity (\eg dropped messages), or highlight good behaviour (\eg fast message delivery) and have built a weight-based scoring function to rank peers according to their behaviour in the network. Every peer keeps a score for every other peer it is connected to and does not share this score with others. Instead, it makes routing and relaying decisions based on its own view of other nodes.
    \item \textbf{The Mitigation Strategies:} Building directly on the properties introduced by the mesh construction and the scoring function, we complement GossipSub with a set of mitigation strategies to protect its operation against malicious activity. Indicatively, these strategies include: controlled mesh maintenance through score-driven participation, scoped flooding for newly published messages and score-based node isolation when malicious activity is detected.
\end{enumerate}

In GossipSub-based systems all nodes start equal and build their profile based on their behaviour. Well-behaved nodes are generally part of the mesh, while nodes with questionable or malicious intentions are progressively excluded from the mesh to the gossip network and then from the network as a whole.

GossipSub is the first of its kind protocol that addresses both the speed and the resilience challenges of permissionless message propagation environments that carry high monetary value. It has been adopted and is currently been integrated as the main transaction and block propagation protocol in the Filecoin network \cite{filecoin}, an incentivized, decentralised storage network, and the Ethereum (ETH2.0) network \cite{eth2}, a decentralised smart contract computation network. Despite carrying utility on top, these systems are primarily financial systems at the transaction and block propagation level and are expected to carry transaction messages potentially worth millions in monetary value. That said, the performance and security properties of GossipSub need to be investigated in detail. That said, it is imperative to detail the performance and security properties of GossipSub, which is the primary focus of this paper.

The contributions of the paper are as follows.
\begin{itemize}[leftmargin=*]
    \item We introduce GossipSub to fill the gap of attack-resistant message propagation protocols for permissionless blockchains. While previous literature has investigated extensively the efficiency of (gossip-based) pubsub protocols, the need for resilience against attacks did not surface due to limited financial incentives from the attacker's point of view.
    \item We build a VM-based testbed environment that supports experiment reproducibility and test the $\sim$16k LOC production code on 5000 containers in AWS. Each container has 1.2 vCPUs and 2 GB RAM. We test \texttt{Sybil:honest} node connection ratios of up to 40:1, although mostly present results with up to 20:1.
    \item We test GossipSub in a wide range of attack scenarios (only a limited part of which is reported here) and validate its resistance against the most primitive of attacks. We find that an average of \$40,000 per month as operational expenditure for coordinated attacks does not constitute a successful attack under any of the metrics derived from the Filecoin or ETH2.0 networks.
\end{itemize}

\section{Background \& Related Work}



In the absence of central control in permission-less, distributed P2P networks there are two components that need particular attention: i) distributed consensus to secure the cryptographic part of the network, and ii) fast and resilient message propagation. Significant research efforts have recently been invested in the first component, that is, in securing permission-less blockchains from the distributed consensus perspective \cite{bano2017consensus}. However, there has been very limited work focusing on the component of fast and secure message propagation \cite{8456488}.




\subsection{The Filecoin Network}

Filecoin is an incentivized, decentralised, P2P storage network. Storage miners contribute storage capacity to the network and get compensated in FIL - the utility token (or cryptocurrency) of the Filecoin network. Filecoin clients are individuals who are interested in storing their data on the Filecoin network. Filecoin users are companies that provide services on top of the Filecoin network, \eg storage and web hosting. Clients operate a \textit{``client node''}, while users operate a \textit{``full node''} in order to participate in the network. Storage miners on the other hand, calculate cryptographic proofs for other miners to verify that they behave honestly and provide valuable service to the network.

Filecoin miners mine on \textit{``(block) tipsets''}. Tipsets are groups of five blocks, which are produced upon every mining round, that is, every 30 seconds. In order to produce the next set of blocks, miners must have received the blocks produced in the previous round \textit{within 6 seconds} after the end of the previous round. Therefore, the deadline of 6 seconds is a hard requirement for the Filecoin blockchain. Failing to meet this deadline will result in a blockchain fork, as miners will not be in sync, \ie they will be mining on different tipsets.

The Filecoin network is currently running in ``testnet'' mode to validate correct operation of all of its components. The size of the ``testnet'' is currently $\sim$300 nodes, with over 6PB of storage committed. The Filecoin mainnet is expected to launch with approximately 1000 nodes in mid 2020 and grow up to 30,000 nodes within the first 12 months. The ratio of miner to full nodes is expected to be in the range of 1:10. Miner nodes are powerful machines in order to calculate the proofs, while full nodes are normal user machines or servers, as needed depending on the application. That said, attacker nodes targeting message propagation are most likely to be full nodes as compared to miner nodes.

The size of transaction messages varies between ~100B and 2KB. The size of Block messages is approximately 1.5KB. In terms of message rates, it is expected that there will be at least two transaction message per second and five block messages per mining round, where each mining round lasts 30 seconds, hence, 10 block messages per minute. At the messaging layer, the pubsub protocol will operate two topics: one for transactions and one for blocks. Reliable, secure and fast message propagation is more important for the blocks topic than for the transactions topic. This is because the worst case scenario for a delayed transaction is for it to be included in a subsequent block, whereas the worst case scenario for a delayed block is a blockchain fork.

\subsection{The ETH2.0 Network}

Ethereum is a smart-contract computation platform (sometimes referred to as a \textit{``programmable blockchain''}) that is operational since 2015. It has been designed to run decentralized computations, in the form of smart contracts, where computations are validated by all nodes in the network. The Ethereum network has evolved into a smart-contract computation platform, used by the vast majority of projects in the area. The network is carrying a total of \$25B in monetary value and is the second biggest cryptocurrency in terms of market cap, after Bitcoin. The first version of Ethereum (ETH1.0) has been using a ``Proof-of-Work'' consensus scheme and a simple message propagation protocol, according to which every node broadcasts every message to \texttt{sqrt(N)} nodes, where \texttt{N} is the size of the network. The second, improved version of Ethereum (ETH2.0) will adopt a ``Proof-of-Stake'' consensus and will also use GossipSub as the message propagation protocol for blocks and transaction messages. The migration to ETH2.0 is expected before the end of 2020.

Unlike Filecoin, the ETH2.0 network has a hard deadline of block propagation at 12 sec. Block messages should in fact be delivered earlier to allow time for validation, cryptographic operations and state management. ETH2.0 will also have many more topics as part of the messaging protocol than Filecoin, starting with up to 70 and with possibility to increase to a few hundred topics that the pubsub protocol will have to handle. The overall size of the network is expected to be in the area of 5k-10k with not all nodes subscribed to all topics. The load in terms of message rates will be larger in ETH2.0 than in the Filecoin network, expected in the area of tens of messages per second per topic.

\subsection{Bitcoin Message Propagation}

The Bitcoin network consists of approximately 60,000 nodes split in public and private nodes. Public nodes have a public IP address, while private ones are those behind NATs (\ie primarily home users). Private nodes open up to 8 outbound connections and no inbound connections (to avoid NAT issues), while public nodes maintain up to 125 inbound connections and 8 outbound connections.
When a node first joins the network, it learns about its first peers from a bootstrap DNS node (these are run by volunteers and pick a random set of peers from the network to introduce to the new node).

The Bitcoin blockchain uses a simple broadcast (\ie flooding) protocol to disseminate transaction and block messages.
The bandwidth requirement for a Bitcoin node to relay transaction messages only, according to the current network size and node connectivity settings is 18GBs/month for private nodes \cite{10.1145/3319535.3354237}. The figure for public nodes who maintain many more connections is a staggering 350GBs/month. As the network grows or as more connections are allowed per node, this figure is expected to increase dramatically, making the Bitcoin network prohibitively expensive to operate.

Several studies have investigated both the topology and the message propagation characteristics of the Bitcoin blockchain \cite{8325269, 7816866, 8369416, 7958588, apostolaki2018sabre}.
In a recent proposal to tackle the redundancy of the Bitcoin propagation protocol, the authors in \cite{10.1145/3319535.3354237} find that almost 90\% of the bandwidth used to propagate transaction messages carry redundant information.
Erlay uses flooding for message propagation between public nodes and two rounds of ``set reconciliation'' with a smart cross-checking technique, called ``minisketch'', between the rest of the nodes in order to synchronise messages across the network.

The security of the Bitcoin broadcast protocol has also been subject to extensive studies \cite{7163021, 7423672}. In \cite{190890},
the authors investigate the bootstrap mechanism of the protocol. The authors in \cite{7816866}
investigate timing attacks on the Bitcoin network in order to infer its topology.




Although the Bitcoin network has not been successfully compromised, the body of work outlined above highlights the importance of security at the network and messaging layer of open, permissionless blockchains.

\subsection{Prior Art in PubSub}

PubSub systems have been deployed extensively in commercial settings in the past and as such have been designed with several requirements in mind \cite{Eugster:2003:MFP:857076.857078}, \cite{dynamoth}, \cite{Malekpour:2011:ERB:2002259.2002287}. Such requirements include reliable delivery in presence of node churn, load-balancing amongst message relay nodes, scalability as networks grow and resource efficiency to avoid excessive duplicate message delivery.

Gossip-based approaches have also received considerable attention, due to the balance they bring between bandwidth consumption and performance guarantees \cite{Rahimian:2011:VGH:2058524.2059505, Baldoni:2007:TTE:1266894.1266898, rappel, Matos:2010:SES:1863145.1863154, 1038579, Gupta:2004:MCP:1045658.1045677, Zhuang:2001:BAS:378344.378347, Setty:2012:PFR:2442626.2442644}.

While these proposals have helped advance the topic of gossip-based pubsub significantly, none of them has been designed with Sybil-resistance in mind. Instead, security in pubsub systems has investigated issues related to \textit{message encryption, digital signatures and access control} \cite{UZUNOV201694, Onica:2016:CPS:2966278.2940296}. As we will show, the performance penalty of conventional pubsub protocols without security protections is so severe in case of attacks that protocol design decisions have virtually no effect. As such, these protocols do not constitute direct points of comparison for GossipSub.
We therefore turn out attention and focus on comparing the behaviour of GossipSub against message propagation protocols used in similar environments, that is, flooding in the Bitcoin and \texttt{sqrt(N)} in the ETH1.0 blockchain.

\section{Attack Types in Permissionless Blockchains}\label{threat-model}

Given the open, permissionless environment of blockchain networks, a message propagation protocol needs to be resistant against the following list of primitive attacks, all of which target delayed message propagation and eventually forking the blockchain.

\textbf{Sybil Attack:} This is the most common form of attack in P2P networks, since creating large numbers of identities is generally cheap resource-wise, unless cryptographic puzzles are included as part of joining the system. In the case of GossipSub, Sybils will attempt to get into the mesh, through a process called grafting, as we will see in the next Section. This is a first step for carrying out all of the following attacks.

\textbf{Eclipse Attack:} This attack can be carried out against a single victim or the whole network. The objective is to silence the victim by refusing to propagate messages from it or to distort its view by delaying message propagation towards it.

\textbf{Censorship Attack:} Sybils seek to establish themselves in the mesh and propagate all messages except those published by the target peer. In contrast to the Eclipse Attack, in the Censorship Attack Sybils appear to behave properly from all vantage points, but hairpin-drop the victim's messages. The objective of the attacker is to censor the target and prevent its messages from reaching the rest of the network.
This attack is difficult to detect by monitoring and scoring peers, as the Sybils build up score by virtue of propagating all other messages.


\textbf{Cold Boot Attack:} In this attack, honest and Sybil nodes join concurrently when the network bootstraps; honest peers attempt to build their mesh, while connecting to both Sybil and honest peers. Since there is no score built up from a warm, honest-only network to protect the mesh, the Sybils manage to largely take over.

The attack can take place in two cases: i) when the network bootstraps with Sybils joining at $t_0$, and ii) when new nodes are joining the network while the network is under attack. Although the first one is quite unlikely to happen, the second is likely to occur as the network grows. 

\textbf{Flash \& Covert Flash Attack:} In the Flash attack, Sybils connect and attack the network at once. In the Covert Flash Attack, Sybils connect to the network but behave properly for some time in order to build up score. Then, they execute a coordinated attack whereby they stop propagating messages altogether in an attempt to completely disrupt the network. The attack is difficult to identify before the attackers turn malicious as they behave properly up to that point and build a good profile.

\section{GossipSub: The Gossiping Mesh Construction}\label{gossipsub-mesh}

GossipSub consists of two tightly-coupled components with each one of them primarily addressing one of the challenges introduced earlier, namely: the \textit{mesh construction} which leverages gossiping for establishing fast, resource-efficient and scalable message propagation (discussed in this Section) and the \textit{score function} for resilience against malicious activity (discussed in the next Section).

\subsection{Peer Discovery}\label{peer-discovery}

Peer discovery can be driven by arbitrary external means and is pushed outside the core functionality of GossipSub. This allows for orthogonal, application-driven development and no external dependencies for the protocol implementation.
Nonetheless, the protocol supports: i) \textit{Peer Exchange}, which allows applications to bootstrap from a known set of bootstrap peers without an external peer discovery mechanism, and ii) \textit{Explicit Peering Agreements}, where the application can specify a list of peers to which nodes should connect when joining.

\subsubsection{Peer Exchange}

This process is supported through either bootstrap nodes or other normal peers.

Bootstrap nodes are maintained by system operators. They have to be stable and operate 
independently of the mesh construction, that is, bootstrap nodes do not maintain connections to the mesh. Bootstrap nodes maintain scores for peers they interact with (we discuss scoring in Sec.~\ref{gossipsub-scoring-function}) and refuse to serve misbehaving peers and/or advertise misbehaving peers' addresses to others. They also participate in propagating gossip. Their role is to facilitate formation and maintenance of the network.

In terms of peer exchange between normal peers, whenever a node is excluded from the mesh, \eg due to over-subscription (see PRUNE message in Section~\ref{building-mesh}), the pruning peer provides it with a list of alternative nodes, which it can use to reconnect, re-build or extend its mesh.

\subsubsection{Explicit Peering Agreements}

With explicit peering, the application can specify a list of peers which nodes should connect to when joining.
For every explicit peer, the router must establish and maintain a bidirectional (reciprocal) connection.
Explicit peering connections exist outside the mesh, which in practice means that every new valid incoming message is forwarded to all of a peer's explicit peers.

\subsection{Building the Mesh}\label{building-mesh}

The \textit{mesh} is the basic building block of the GossipSub protocol. Each node maintains a list of peers with which it is directly connected with bidirectional, reciprocal links, forming its \textit{local mesh}.

\begin{figure}[h]
\centering
\includegraphics[width=0.5\textwidth]{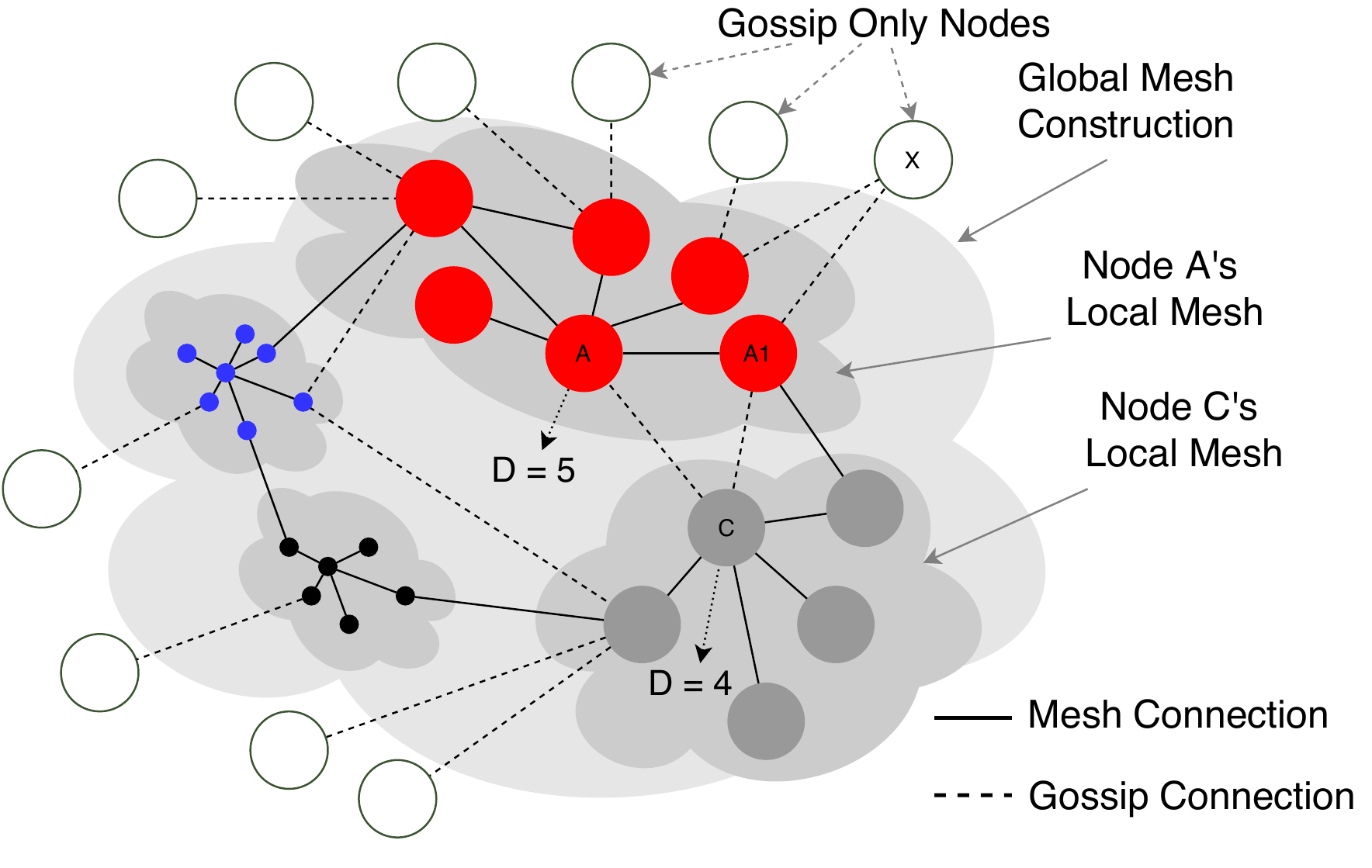}
\caption{The GossipSub Mesh Construction}
\label{fig:gossip-mesh}
\end{figure}

The local mesh of each node is the group of nodes that connects it to the \textit{global mesh}. Messages propagate through their first-hop connections to reach neighbour local meshes and propagate further. 
In Fig.~\ref{fig:gossip-mesh} we present the structure and interactions between nodes in GossipSub.

When a node wants to either publish or relay a message it has received, it broadcasts the message to its \textit{local mesh only}. The recipients of that message that have built and maintain their own \textit{local-meshes} will propagate it further through the construction. Note that it is not necessary for every participating node to have an established \textit{local mesh} as nodes can issue direct requests through gossip for messages that they have not received (we describe the details of the gossip process in Section~\ref{gossip-section}).

The number of peers that a node maintains a direct connection with as part of its \textit{local mesh}, \texttt{D}, is the desired degree of the network per topic and therefore, the amplification factor of the protocol.

The overlay is initially constructed according to the peer discovery mechanisms discussed in Section~\ref{peer-discovery}. The parameters of GossipSub related to the mesh construction are: \texttt{D}, which is the target degree, and two relaxed degree parameters \texttt{D\_low} and \texttt{D\_high}, representing admissible mesh degree bounds. The goal of the degree parameters is to introduce elasticity and curtail excessive flapping. When the direct connections in a node's mesh exceed \texttt{D\_high}, the node is \textit{over-subscribed} and needs to exclude (PRUNE) some nodes from its mesh,  while when the number of connections falls below \texttt{D\_low}, the node is \textit{under-subscribed} and needs to include (GRAFT) new nodes in its mesh.

\begin{figure}[h]
\centering
\includegraphics[width=0.3\textwidth]{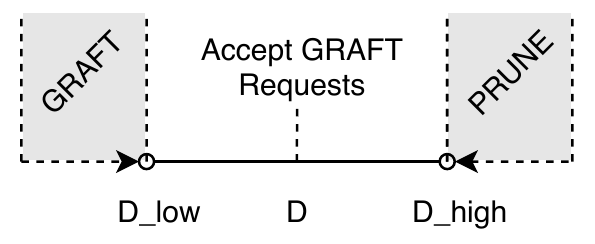}
\caption{Balancing the Mesh}
\label{fig:d-values}
\end{figure}

Higher values for the \texttt{D\_*} values translate to lower message propagation latency as the \textit{global mesh} is more densely connected. However, better connectivity comes with higher bandwidth cost as messages are directly forwarded to more peers. For the purposes of this study, we have set \texttt{D=8, D\_low=6, D\_high=12} in order to preserve low-fan out and therefore, reduce bandwidth requirements, but at the same time achieve timely message propagation.
We discuss tradeoffs in the next subsection and carry out a detailed sensitivity analysis to justify our choices in Section~\ref{sec:sensitivity-analysis}.

This set of parameters is later extended to include: i) \texttt{D\_score} to rank peers in the mesh, according to their behaviour, as observed by the scoring function, and ii) \texttt{D\_out} to set a minimum number of outbound connections per peer. We discuss these later in Section~\ref{mitigation-strategies}.



The protocol defines five control messages and a \textit{heartbeat}:

\begin{itemize}[leftmargin=*]
\item GRAFT a mesh link: this notifies the peer that it has been added to the local mesh of the grafting node.
\item PRUNE a mesh link: this notifies the peer that it has been removed from the local mesh of the pruning peer.
\item PRUNE-Peer Exchange (PX): the pruning peer sends a list of peer IDs to the pruned peer to help it connect to alternative peers and expand its mesh. There is a backoff timer associated with pruning to avoid re-grafting to recent peers for a period of time (set to 1min).
\item IHAVE: gossip; this notifies the peer that the following messages were recently seen and are available on request.
\item IWANT: gossip; request transmission of messages announced in an IHAVE message.
\end{itemize}

The protocol also implements two simple messages to join and leave a topic, \texttt{JOIN[topic]} and \texttt{LEAVE[topic]}, which are implemented through a \texttt{GRAFT[topic]} and \texttt{LEAVE[topic]} message, respectively.

Finally, the router periodically runs a \textit{heartbeat} procedure, which is responsible for maintaining the mesh and emitting gossip.
The value of the heartbeat is set to 1\,s in our current implementation. As we will discuss later in Section~\ref{mitigation-strategies}, mesh maintenance is responsible for clearing out badly-behaving nodes from the mesh.


\subsection{The Gossip Factor}\label{gossip-section}


Gossip messages propagate metadata throughout the network. The metadata can be arbitrary, but as a baseline we include the message IDs that the router emitting the gossip has seen in the last three seconds.

Gossip is emitted to a random subset of peers that \textit{may or may not be part of the mesh}. It is important to note that gossiping does not happen exclusively between nodes outside the mesh to/from nodes inside the mesh as shown in Fig.~\ref{fig:gossip-mesh}. In other words, these are not two mutually exclusive groups. All peers emit gossip and direct it to random peers through their connection lists. For instance, Peer A1 in Fig.~\ref{fig:gossip-mesh} can gossip both to Node X, which is currently not part of the mesh and to C, which \textit{is} part of the mesh. It is worth highlighting though that nodes connected to the mesh are more likely to have received messages before the gossip arrives at the next heartbeat (\ie $\sim$1\,s later).


Gossip is emitted on every heartbeat according to a window-based mcache shifting procedure to cover all messages seen by a peer. In our current implementation, GossipSub is running three rounds of gossip upon three consecutive heartbeats. In every round, peers choose only a subset of the peers they know to emit gossip to. Therefore, not all peers will ``hear'' about a message from the first gossip round and hence, a second and third round is needed in order to guarantee that all peers have received all messages. We discuss specific settings for gossip dissemination as well as the choice of three gossip rounds later, in Section~\ref{sec:adaptive-gossip}. 
\\[0.3cm]
Summarising, \textit{GossipSub is a blend of mesh-restricted eager push for data and gossip-based lazy-pull for metadata.} The combination of those techniques guarantees responsiveness to network conditions, but most importantly fast propagation as well as scalability in terms of traffic volume.

\begin{figure}[h]
\centering
\includegraphics[width=0.3\textwidth]{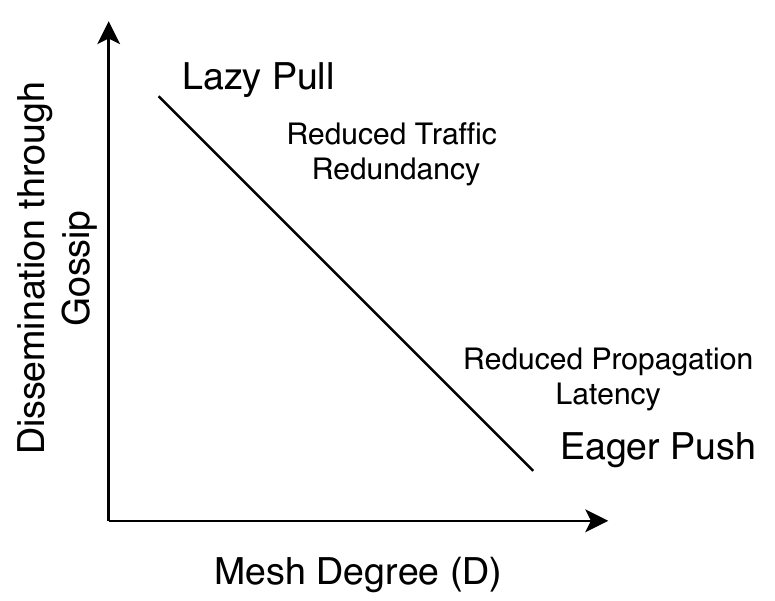}
\caption{GossipSub Tradeoffs}
\label{fig:tradeoffs}
\end{figure}

Eager push is more bandwidth demanding but results in faster message propagation. There is a clear tradeoff between the degree of the mesh (and, therefore, the volume of eager push traffic) and the amount of gossip that needs to be emitted in order to cover the entirety of the network. In Fig.~\ref{fig:tradeoffs} we present the relationship between these two components of GossipSub. We present a thorough sensitivity analysis for the degree of the mesh (\ie the value of \texttt{D} and its degree parameters) in Section~\ref{sec:sensitivity-analysis}.

\section{GossipSub: The Score Function}\label{gossipsub-scoring-function}

The score function is used as a performance monitoring mechanism to identify and remove badly-behaving nodes from the mesh. Each peer in the mesh is keeping a score for every other peer\footnote{The overhead of keeping state for other peers is minimal, in the area of $~0.5$KB per observed peer.} that it directly interacts with\footnote{Every peer keeps only a partial view of the network; there is no peer with the full view of the network or the mesh.} but \textit{scores are not shared between peers}. Sharing one's view of another peer with others would effectively result in a reputation system. Reputation systems come with their own set of challenges, \eg colluding attacks, which we have left for future investigation. The scoring mechanism is plugged into various GossipSub algorithms such that peers that behave correctly attain positive score, while those that misbehave are penalised with negative score and are eventually removed from the mesh.

\subsection{Building the Score Function}

We have designed a generic and modular function to incorporate all the important parameters through which we can identify honest or malicious behaviour, in order to reward or penalise peers accordingly. The modularity of the score function allows for addition of more parameters depending on application-specific deployments. The score function is a weighted sum of parameters with the weights being set according to environment conditions. Some parameters are specific to topics, while others apply across all topics that the observing peer is subscribed to.

\begin{equation}
    Score(peer) = TC(\sum_{n=1}^{4} w_n(t_i)*P_n(t_i)) + w_5*P_5 + w_6*P_6
\end{equation}
where $TC$ is the \textit{TopicCap} and $t_i$ is the topic weight for each topic where per topic parameters apply. Note that for the purposes of this study, we experiment with one topic only and in particular we consider the Block propagation topic of the Filecoin and ETH2.0 networks.

The topic parameters are implemented using counters maintained internally by the router whenever an event of interest occurs. The parameters are defined as follows.

\begin{itemize}[leftmargin=*]
    \item $P_1$: \textbf{Time in Mesh} (for a topic). This is the time a peer has been in the mesh, capped to a small value and mixed with a small positive weight. $P_1$ is intended to mitigate flash attacks and boost peers already in the mesh.
    \item $P_2$: \textbf{First Message Deliveries} (for a topic). This is the number of messages first delivered by the peer in the topic, mixed with a positive weight. $P_2$ is intended to reward peers who act fast on relaying messages.
    \item $P_{3a}$: \textbf{Mesh Message Delivery Rate} (for a topic). This parameter is a threshold for the expected message delivery rate within the local mesh. If the number of deliveries is above the threshold, then the value is 0. If the number is below the threshold, then the value of the parameter is the square of the deficit. $P_{3a}$ is intended to penalize peers in the mesh who are likely to silently be dropping messages so that they are removed from the mesh. The parameter is mixed with a negative weight and is capped. The implementation of $P_{3a}$ has to include a configurable time window within which message delivery is considered valid. The time window has to be kept in order to avoid a peer building score by simply replaying back messages received by a particular peer.
    \item $P_{3b}$: \textbf{Mesh Message Delivery Failures} (for a topic). $P_{3b}$ counts the number of mesh message delivery failures. Whenever a peer is pruned with a negative score, the parameter is augmented by the rate deficit at the time of prune. $P_{3b}$ is intended to keep history of prunes so that a peer which was pruned because of under-delivery cannot quickly re-graft onto the mesh. The parameter is mixed with negative weight and is uncapped, in order to create stickiness of negative score for as long as the peer misbehaves.
    \item $P_4$: \textbf{Invalid Messages} (for a topic). The $P_4$ parameter is intended to penalize peers who transmit invalid messages, according to application-specific validation rules. It is mixed with a negative weight and is uncapped, again, in order to increasingly penalise peers for as long as they send invalid messages.
    \item $P_5$: \textbf{Application-Specific score.} $P_5$ is the score component assigned to the peer using application-specific logic. The weight is positive, but the parameter itself has an arbitrary real value, so that the application can signal misbehaviour with a negative score or gate peers before an application-specific handshake is completed. One example of an application-specific score is staked participation \cite{pop-icbc-2018, gupta-reputation-currency}.
    \item $P_6$: \textbf{IP Address Collocation Factor.}
    $P_6$ is intended to make it difficult to carry out Sybil attacks by using a small number of IP addresses.
    If the number of peers connecting from the same IP exceeds an application-specific threshold, then the value of $P_6$ is the square of the surplus, otherwise it is 0. The parameter is mixed with a negative weight.
\end{itemize}

The combination of score function parameters have been designed such that are difficult to manipulate. For example, staying connected in the mesh for a long time (to build up $P_1$) and behaving properly (\ie forwarding all messages) incurs operational costs to the attacker. In Sec.~\ref{sec:bandwidth-requirements} we show that the cost of an attack can potentially be as high as \$40,000/month. On the other hand, staying connected but not forwarding traffic, in order to save bandwidth costs will trigger negative scoring for $P_2$ and $P_{3a}$.


\subsection{Topic Value Decay}

The counters for all of these function parameters decay periodically so that their values are not consistently increasing (or decreasing) and ensure that a large positive or negative score is not sticky for the lifetime of the peer. The decay interval is configurable by the application, with shorter intervals resulting in faster decay.

The decay factor and interval together determine the absolute rate of decay for each parameter. For example, a decay interval of 1 second and a decay factor of 0.90, will cause a decrease of 10\% every second.

\section{GossipSub Mitigation Strategies}\label{mitigation-strategies}

We discuss the mitigation strategies that we have incorporated into GossipSub in decreasing level of effectiveness and complexity, all of which are tailor-made to suit the main building blocks and features of the protocol. We map mitigation strategies to attacks in Table~\ref{table:mitigation-strategies}.

\subsection{Controlled Mesh Maintenance}\label{sec:mesh-maintenance}

The GRAFT mechanism of GossipSub creates an attack vector, as malicious nodes can create multiple Sybil identities and send GRAFT messages to honest peers. As more Sybil nodes attempt GRAFT requests, honest peers go over the \texttt{D\_{high}} threshold (discussed in Section~\ref{building-mesh}). If peers PRUNE their connections \textit{randomly} in order to keep between \texttt{D\_{low}} and \texttt{D\_{high}} (Fig.~\ref{fig:d-values}), they will (statistically) prune honest peers in favour of malicious ones, especially as the \texttt{sybil:honest} node ratio increases. The situation results in malicious nodes controlling most or all of the connections in honest nodes' local meshes.

As an immediate mitigation strategy, GossipSub is pruning peers \textit{selectively}. The score that peers keep for their neighbours enables them to make more informed pruning decisions. In particular, when pruning because of oversubscription, the peer keeps the best scoring peers from its existing mesh and  selects the remaining peers to GRAFT at random from its list of known/seen peers.
This protects the mesh from takeover attacks and ensures that the best scoring peers are kept in the mesh. At the same time, GossipSub allows for grafting of a few new peers, so that the protocol is responsive to new peers joining the network.

We realise this concept through two extra mesh maintenance variables.

\subsubsection{Index of best performing peers, \texttt{D\_score}}
This is an index of the best-performing existing peers in the node's local mesh.
In the current implementation we set \texttt{D\_score=6}. Recall that \texttt{D=8, D\_{high}=12, D\_{low}}=6. 

\subsubsection{Outbound Mesh Connection Quotas, \texttt{D\_out}}

We are augmenting the mesh construction with connection directionality (\ie inbound vs outbound connections). Outbound connections to honest peers are important in an adversarial environment in order for a node that is under attack to avoid being silenced. From the Sybils' point of view, it is easier to join the network and attempt to occupy all incoming connections to the target node by simply opening multiple connections to it. Instead, it is more difficult to attract outbound connections, which are initiated by the honest node itself.
\texttt{D\_out} is adjusted during the mesh maintenance so that it is lower than \texttt{D\_low}, but never exceeds \texttt{D/2}.


Summarising, the mesh maintenance process is run on every heartbeat (\ie every 1\,s) to: i) prune peers with negative score from the node's local mesh, ii) choose only peers with positive scores to graft to, in case of under-subscription, and iii) always maintain at least \texttt{D\_out} outgoing connections.
\\[0.15cm]
\textbf{Mitigates:} \textit{Sybil, flash and eclipse attacks}, according to which an attacker is spinning up a large number of Sybils to take over the mesh.

\subsection{Opportunistic Grafting}\label{sec:oppgraft}

In cases of significant rates of churn of honest nodes, or a successful take over of a peer's mesh, an honest peer can get stuck with a mesh of poorly performing peers. When this happens, the router will normally react through mesh failure penalties ($P_{3a}, P_4, P_6$) in Section~\ref{gossipsub-scoring-function}), but this reaction time can be slow: the peers selected to replace the negative scoring peers are chosen at random among the non-negative scoring peers, which may result in multiple rounds of selections from a Sybil-poisoned pool before the mesh becomes healthy again. Furthermore, the population of Sybils may be so large that the sticky mesh failure penalties completely decay before any good peers are selected, thus making Sybils re-eligible for grafting.

In order to recover from these extreme conditions, we have introduced an \textit{``opportunistic grafting''} mechanism, according to which peers periodically check the median score of other peers in their mesh against a threshold. This threshold is a preset positive real[/integer], defined based on experimental evaluation of a healthy network. If the median score is below the threshold, the peer opportunistically grafts extra peers with score above the median in the mesh. This improves an under-performing mesh by introducing good-scoring peers that may have been gossiping to honest nodes.


Opportunistic grafting is applied periodically with the recommended period being set to 1 min, while 
the number of peers that are opportunistically grafted is controlled by the application. For instance, it may be desirable to graft more peers if the application has configured a larger mesh than the default parameters.
\\[0.15cm]
\textbf{Mitigates:} Successful mesh take-over and therefore, a poisoned-pool mesh, seen in the Cold-Boot and Covert Flash Attacks.

\subsection{Flood Publishing}\label{sec:flood-publishing}


We are introducing \textit{``Flood Publishing''}, according to which every newly published message is sent to all known peers with a positive score that are subscribed to the topic.
This applies regardless of whether the publisher is subscribed to the topic.

As is the case with flooding more generally, this behaviour reduces message propagation latency at the cost of increased bandwidth requirement. However, the fact that in GossipSub we use flooding for the publisher node only and not for the whole network strikes the right balance between avoiding overloading the network, achieving good propagation latency and achieving good protection against eclipse attacks.
\\[0.15cm]
\textbf{Mitigates:} Flood publishing is very effective in countering all identified attacks where Sybils attempt to eclipse or censor a specific target or the whole network in either warm or cold conditions.

\subsection{Adaptive Gossip Dissemination}\label{sec:adaptive-gossip}

In order to counter against varying number of Sybils in the network, we have decided to dynamically adjust the gossip that the protocol emits \textit{based on the peer's known connections}.
The parameter controlling the emission of gossip is called \textit{``gossip factor''} and is set to 0.25 (25\%) for the purposes of the present study. This means that in each round every peer is emitting gossip to 25\% of its known peers that are not part of its local mesh. The specific setting of 0.25 for the \textit{gossip factor} was chosen so that every node has at least 50\% probability of receiving a message through gossip.\footnote{Given $gossip\ factor = 0.25$, the probability of a node \textit{not} receiving a message through gossip is $0.75^{rounds}$, or 0.42 for the three gossip rounds, which leaves us with a probability of 0.58 of receiving a message through gossip. Note that this figure is independent of the Sybil to honest node ratio.}
\\[0.15cm]
\textbf{Mitigates:} Enhances the resistance of the protocol against eclipse attacks, in case of significant number of Sybils.

\subsection{Backoff on PRUNE}\label{sec:backoff-prune}

According to GossipSub's mesh construction rules, when a node is oversubscribed, in terms of number of nodes in its local mesh, it prunes peers to bring its mesh back to normal size (below \texttt{D\_high}).
This functionality is extended to add a \textit{backoff} period before a pruned peer can attempt to re-graft. This action helps preserve the mesh from Sybil nodes that attempt continuous GRAFTs to take over the mesh of the node. The recommended duration for the backoff is 1 min.
\\[0.15cm]
\textbf{Mitigates:} GRAFT spam attacks and generally keeping malicious nodes away from the mesh.

\begin{table*}[t]
\centering
\begin{tabular}{|p{6cm}||P{1.2cm}|P{1.2cm}|P{1.2cm}|P{1.2cm}|P{1.2cm}|}
 \hline
 \multicolumn{6}{|c|}{GossipSub Attack \& Mitigation Strategy Summary} \\
 \hline
\diagbox[innerwidth=6cm]{Mitigation}{Attack} & Sybil & Eclipse & Censor  & Cold Boot & Covert Flash\\
 \hline
 Backoff on PRUNE (Sec.~\ref{sec:backoff-prune}) & \checkmark &  &  & & \\
 Adaptive Gossip Dissemination (Sec.~\ref{sec:adaptive-gossip}) & \checkmark & \checkmark  & & & \\
 Controlled Mesh Maintenance (Sec.~\ref{sec:mesh-maintenance}) & \checkmark & \checkmark & & \checkmark & \checkmark \\
  Opportunistic Grafting (Sec.~\ref{sec:oppgraft}) & \checkmark & \checkmark & \checkmark & \checkmark & \checkmark \\
  Flood Publishing (Sec.~\ref{sec:flood-publishing} & \checkmark & \checkmark & \checkmark & \checkmark & \checkmark \\
\hline
\end{tabular}
\caption{GossipSub Attacks vs Mitigation Strategies}
\label{table:mitigation-strategies}
\end{table*}
\section{Infrastructure Setup \& Evaluation Methodology}\label{infra}

\subsection{The Testground Environment}

In order to evaluate the performance of GossipSub under realistic conditions, we have built Testground, a platform for testing, benchmarking, and simulating distributed and peer-to-peer systems at scale. Testground is designed to support reproducibility of results, it is platform-agnostic, but most importantly is able to scale to any number of instances. With Testground tests can run either as executables or docker containers locally, or in Kubernetes-based cloud environments. In Testground VMs are instantiated and connected according to configuration files written separately. We provide a detailed description of our test environment in Section~\ref{sec:testground-description} and we make all of our test plans and configuration files publicly available for reproducibility.\footnote{Hidden for double-blind review.} The codebase for GossipSub is approximately 16000 LOC\footnote{Publicly available at: \url{https://github.com/libp2p/go-libp2p-pubsub}}, with about half of it for event tracing and tests.



We have run our evaluation using Testground Kubernetes clusters on AWS and allocate 1.2 vCPUs and 2 GBs of memory per container node.

\subsection{Evaluation Methodology}

We tested GossipSub in many more attack scenarios than discussed in Section~\ref{threat-model}, but can only present a subset of them due to space limitations. In particular, we present results for the following three attacks: i) Network-Wide Eclipse Attack, ii) Cold Boot Attack, and iii) Covert Flash Attack. This set of attacks covers  the most challenging scenarios and provides a detailed insight into the behaviour of the protocol.

We compare GossipSub against:
\begin{enumerate}[leftmargin=*]
    \item \textbf{Bitcoin's Flooding/Broadcast protocol.} We emulate the performance of the Bitcoin blockchain at the network layer by assuming only public nodes, which open 133 connections in total (125 inbound and 8 outbound).
    \item \textbf{ETH1.0's \texttt{sqrt(N)}-based pubsub protocol}, according to which newly published messages are forwarded to the square root of the number of nodes in the network. For our tests in this study, where we assume 1000 honest nodes, the degree of the \texttt{sqrt(N)} protocol equals 32, which is the setting we are using.
    \item \textbf{Plain GossipSub} (occasionally denoted as pGM) for a version of GossipSub without the scoring function or the attack mitigation strategies.
\end{enumerate}

We are using 2) and 3) above as benchmark protocols to approximate the performance of traditional pubsub protocols. We highlight that although this is a simplistic approach to the comparison against the vast related literature, here we want to focus on the mitigation properties of GossipSub in adversarial environments. As we show in the next Section, the attacks investigated have such severe impact on protocol performance that non-security design choices do not actually affect performance much.

We believe that this set of protocols captures the whole picture for message propagation in open, permissionless blockchains, according to current practices and the most popular blockchain networks in operation today.

\subsection{Evaluation Setup \& Metrics}

Unless otherwise mentioned, the following settings apply to our test scenarios.
We setup 1000 honest nodes (100 publishers and 900 full nodes) and 4000 Sybils targeting every honest node in the network. We assume that Sybils can establish 100 connections each, while honest nodes only up to 20. This brings the \texttt{Sybil:honest} connection ratio to 20:1.\footnote{We have experimented with smaller as well as bigger connection ratios and we confirm that performance trends remain as reported here.} We have chosen these settings for the maximum number of Sybil and honest node connections in order to evaluate a worst-case scenario. In reality, honest nodes will be able to open many more connections, subject to bandwidth and CPU resources.



We focus our attention on the Block topic in terms of message delivery deadlines, where the Filecoin and ETH2.0 networks have hard global delivery deadlines at 6\,s and 12\,s, respectively, after message generation. However, in order to stress-test the system we apply much higher message rates and experiment with 120 messages/sec.\footnote{We have tested the performance of all protocols with message rates up to 1000 msgs/sec and confirm that performance follows similar trends.}

The primary metrics of successful attack mitigation are: i) Cumulative Distribution Function (CDF) \textit{lower than 6\,s} for the Filecoin network and 12\,s for ETH2.0, and ii) the percentage of message loss. We report early on that we have not observed message loss for GossipSub in any of the tests we have run. We report message loss for the other protocols where applicable. We measure the Probability Distribution Function and present the $99^{th}$ percentile of the global propagation latency (denoted as ``p99'') and also capture the score profile of honest and attacker peers in order to get an understanding of the protocol's behaviour. Last, but not least, we monitor the bandwidth cost of message propagation in terms of the number of duplicate message delivery.
\section{Performance Evaluation}\label{perf-eval}

\subsection{Network-wide Eclipse Attack}

Given GossipSub's mesh construction properties, the Eclipse attack is executed by having a large number of Sybils continuously attempting to graft onto the mesh. They Sybils are targeting all nodes in the network and attempt to occupy as many connections of each node as possible. Once they do, they drop all messages to and from the nodes they have managed to connect to.

The attack takes place in a warm network, where honest nodes join the network and form meshes at $t_0 = 0sec$, while attackers are introduced at $t_1 = 60 sec$ into the test. We have confirmed that this is a long-enough interval for honest nodes to setup their local meshes and build a score profile by varying the warm-up time.

\begin{figure}[h]
\centering
\includegraphics[width=0.3\textwidth]{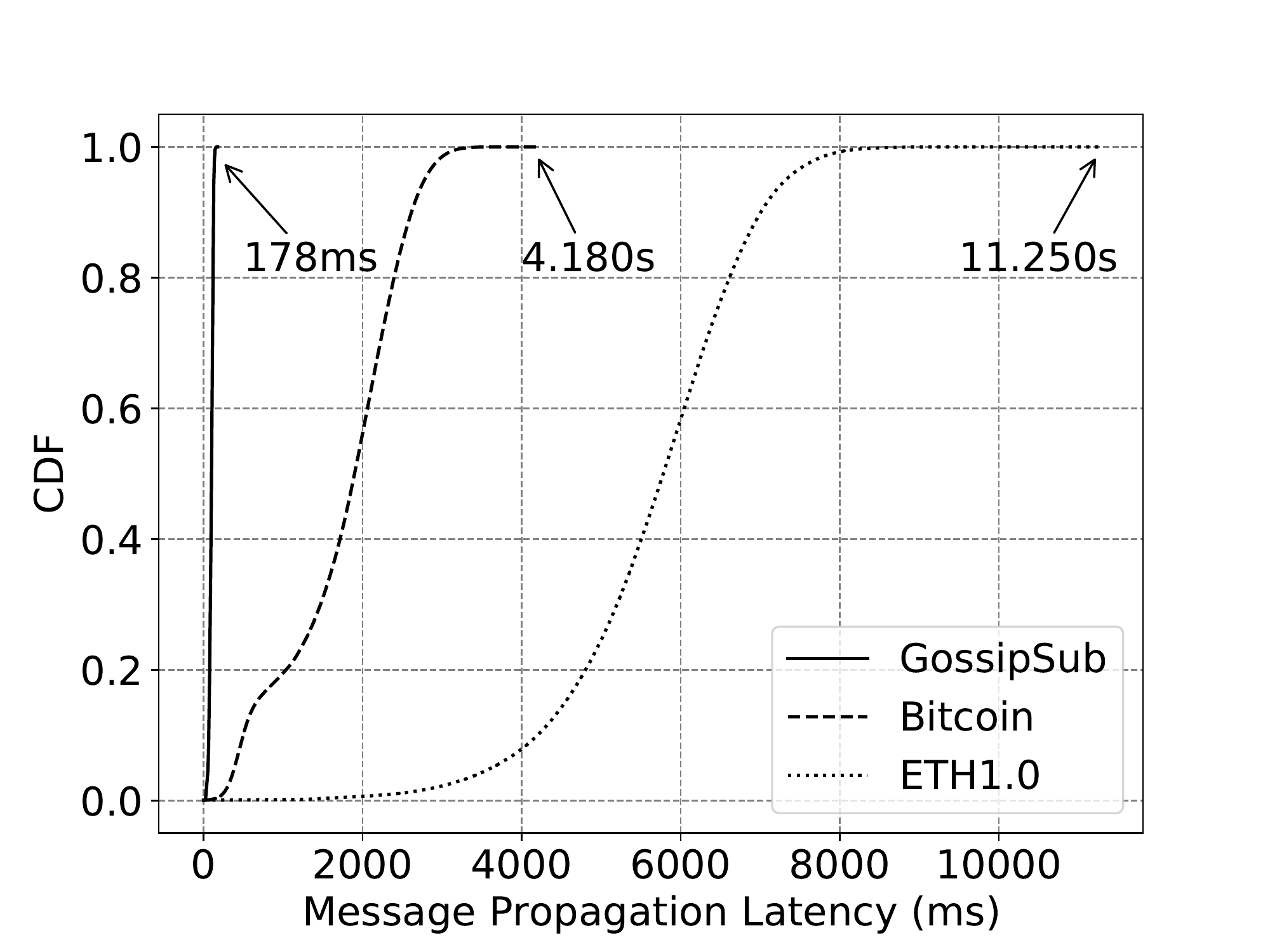}
\caption{Network-Wide Eclipse Attack}
\label{fig:eclipse-net-merged-cdf}
\end{figure}

Our results show that \textit{plain GossipSub} is completely devastated by the attack. The system starts losing messages immediately with the latency of delivered packets reaching up to 50secs (not presented here for clarity of Fig.~\ref{fig:eclipse-net-merged-cdf}) and $\sim10\%$ message loss.
The results indicate a successful attack in taking over the mesh with successful message propagation happening almost purely through gossip.

In contrast, GossipSub is totally immune to the attack. The attackers only manage to occupy a few slots in the mesh, which does not make any dent to the delivery rate. The maximum message propagation latency is at 178ms (see Fig.~\ref{fig:eclipse-net-merged-cdf}) with the p99 at 141 ms.

These values are identical to the ``Baseline Scenario'' results that we present in Section~\ref{sec:sensitivity-analysis}, where there are no attacker nodes. This behaviour indicates that most attacker nodes do not even make it to the mesh, given that the attack happens on a warm network.

In particular, selective score-driven peer selection ensures that nodes are accepting just a few new/low-scoring nodes in the mesh, keeping a portion of their connections grafted to their old, high-scoring peers. Therefore, an army of Sybils attacking a warm node will only succeed in occupying a minority of slots in the mesh. In this particular test, malicious nodes occupy between 2 and 4 connections on average in nodes' local meshes, leaving enough connections to honest nodes to propagate messages quickly through the network.

ETH1.0's \texttt{sqrt(N)}-degree pubsub protocol completely fails to resist the attack and results in total message propagation of more than 11s with the p99 at $\sim$8s. It turns out that even Bitcoin's simple broadcast protocol with 133 total connections per node is also severely affected with a CDF above 4s (Fig.~\ref{fig:eclipse-net-merged-cdf}) and p99 at 3s - roughly three times faster than ETH1.0's \texttt{sqrt(N)}, \ie approximately equal to the fan-out difference of the two protocols.\footnote{Recall that for a network of 1000 nodes, ETH1.0's protocol is forwarding maintaining $\sim32$ connections.}

\subsection{Cold Boot Attack}\label{sec:cold-boot}


In this attack, honest and Sybil nodes join the network concurrently. In fact, in order to test the worst case scenario, we let Sybil nodes connect first and form meshes between them. Honest peers join 2 min into the test and attempt to build their mesh in a Sybil-dominated network. As a result most of the \texttt{D} connections of honest nodes are opened towards Sybil nodes. Sybils dominate the mesh and carry out an eclipse attack, where they drop both incoming and outgoing messages.



\begin{figure}
     \centering
     \begin{subfigure}
         \centering
         \includegraphics[width=0.225\textwidth]{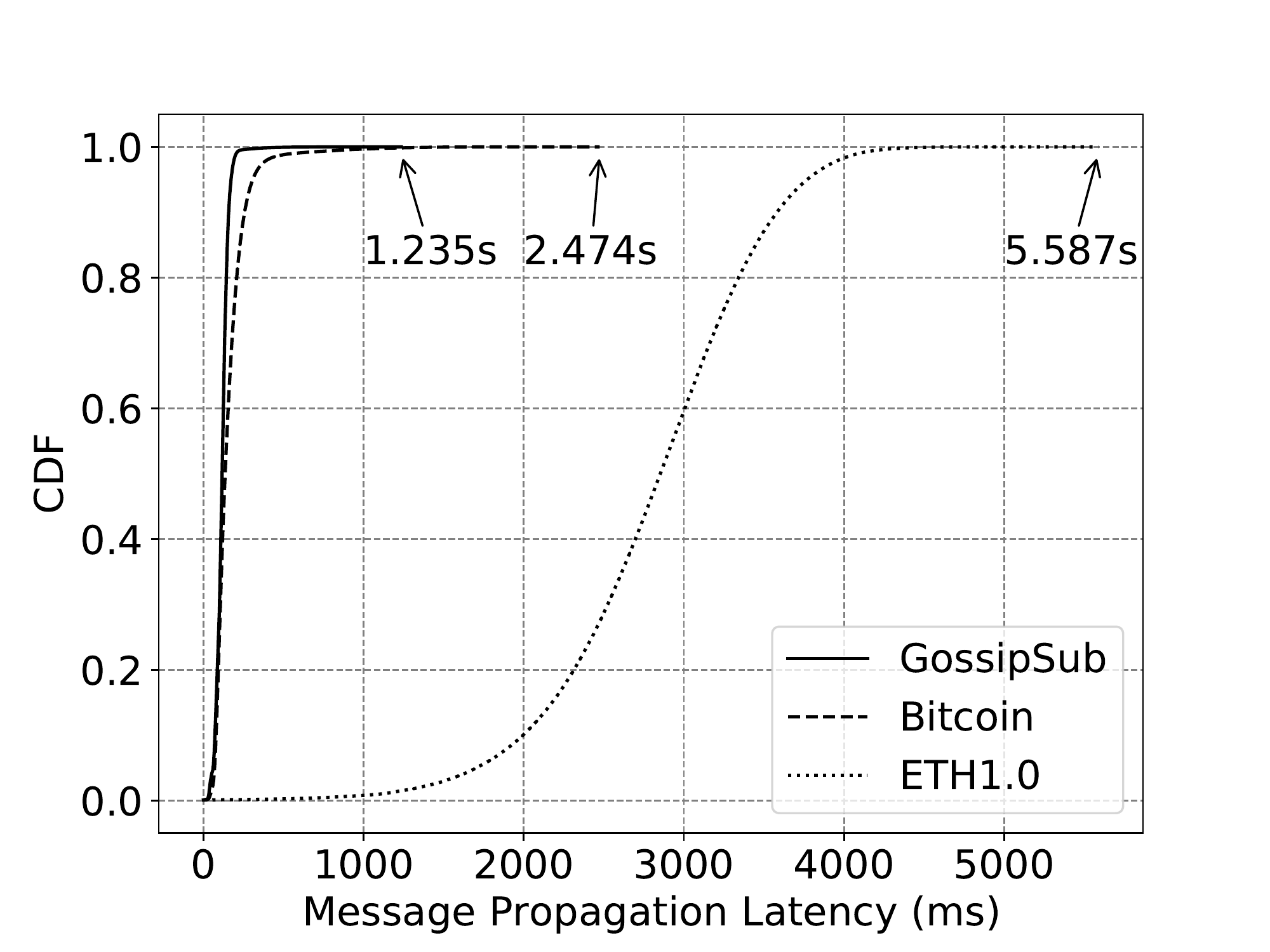}
     \end{subfigure}
     \begin{subfigure}
         \centering
         \includegraphics[width=0.225\textwidth]{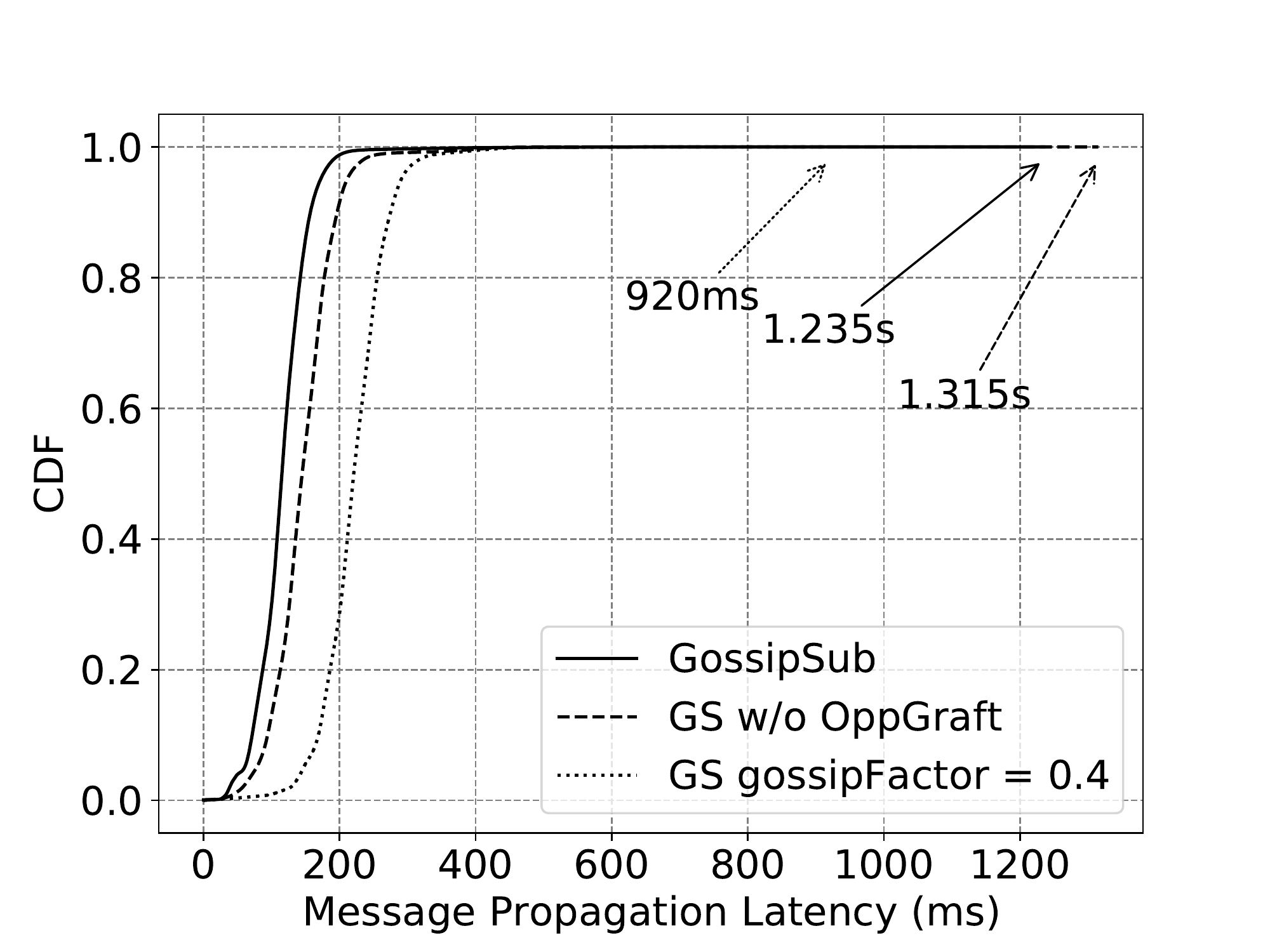}
     \end{subfigure}
\vspace{-0.5cm}
\caption{Cold Boot Attack}
\label{fig:cold-boot-merged-cdf}
\end{figure}

Plain GossipSub fails to respond to the attack. It not only delivers messages with prohibitively long delay (up to 17\,s and p99 at 10\,s) but also fails to propagate $\sim$4\% of messages.
GossipSub achieves 100\% message propagation and is only slightly affected by the attack with some messages delayed for up to 1.2\,s as shown Fig.~\ref{fig:cold-boot-merged-cdf} - well within the 6\,s window. Bitcoin's broadcast protocol increases message propagation latency by 100\% compared to GossipSub to 2.4\,s, but still achieves somewhat acceptable performance, being able to meet the 6\,s deadline of Filecoin and ETH2.0 networks. Note, however, that this comes at a severe bandwidth cost: Bitcoin's broadcast protocol sends an order of magnitude more duplicate messages, 1.8M, compared to GossipSub's 153k. We argue that this is an enormous cost, which although makes an attack against the network more expensive in terms of bandwidth requirement for malicious nodes, it also increases significantly the cost of operating honest nodes.\footnote{We discuss bandwidth requirements in more detail in Section~\ref{sec:bandwidth-requirements}.} ETH1.0's \texttt{sqrt(N)}-based protocol fails to recover from the attack with most of its nodes remaining primarily connected to Sybils. The protocol is failing to meet the deadline with the CDF of global propagation latency at almost 6\,s again shown in Fig.~\ref{fig:cold-boot-merged-cdf} (left) and p99 at 4\,s.


\begin{figure}[h]
\centering
\includegraphics[width=0.33\textwidth]{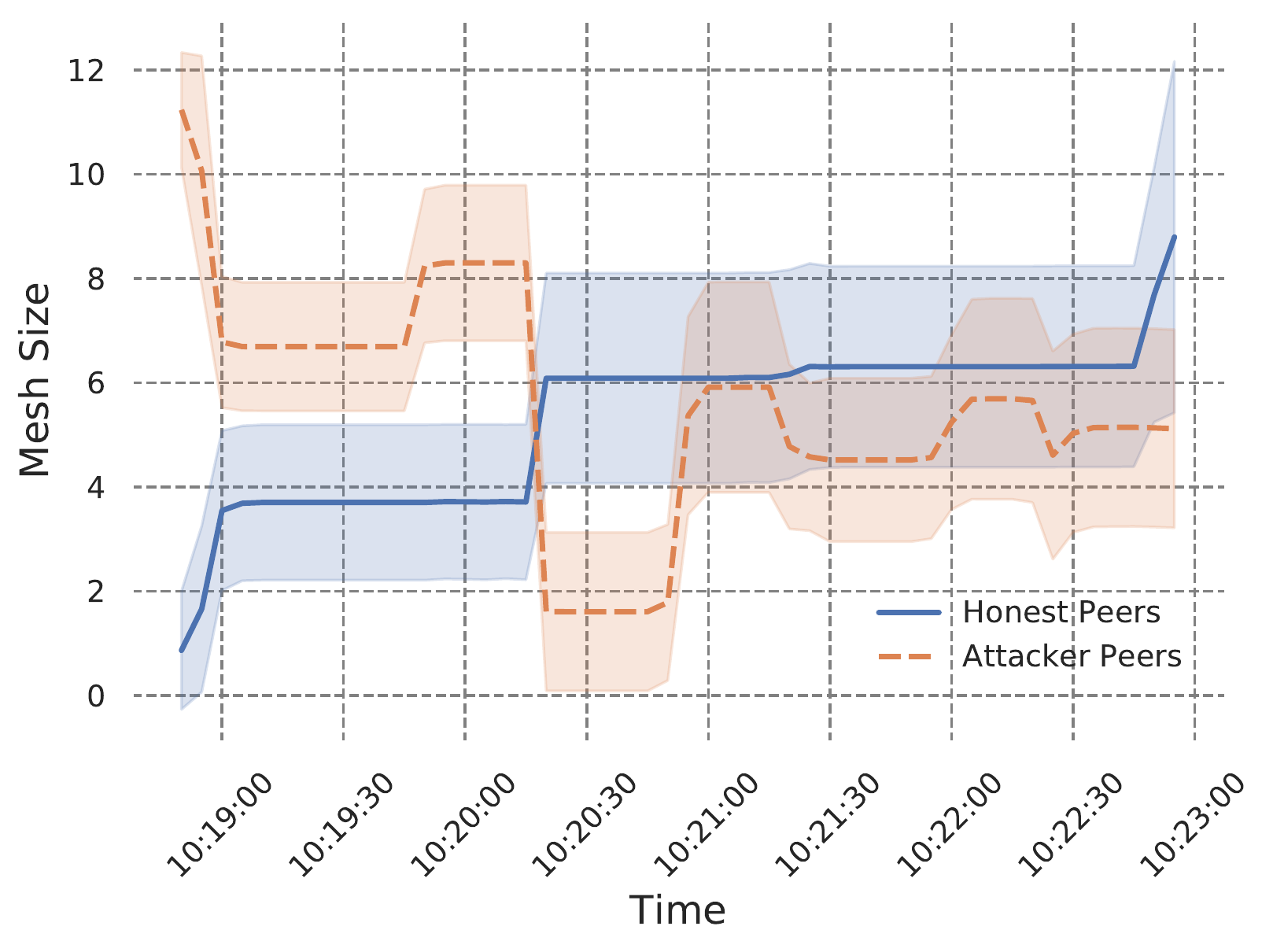}
\caption{Mesh Status - Cold Boot Attack}
\label{fig:mesh-status-cold-boot}
\end{figure}

In Fig. \ref{fig:mesh-status-cold-boot} we present the state of the mesh throughout the duration of the experiment. Given the experiment setup, the mesh is initially poisoned as Sybils join the network before honest nodes and dominate mesh connections. It is remarkable to see that GossipSub manages to recover the mesh within 90 heartbeats, \ie $\sim$1.5 m. Despite the Sybil-dominance duration, GossipSub manages to neither lose messages nor delay message propagation prohibitively.

We present results from a test without opportunistic grafting, and also include a test with the gossip factor increased to 0.4 from its default value of 0.25 (Section~\ref{sec:adaptive-gossip}). With the default settings (\ie \texttt{gossipFactor = 0.25} and \texttt{OppGraft} enabled) the p99 is 205\,ms, while disabling opportunistic grafting increases the p99 to 269\,ms. The CDF for these cases, as well as for the case when we increase the \texttt{gossipFactor = 0.4} is shown in Fig.~\ref{fig:cold-boot-merged-cdf} (right). We see that increased gossip dissemination reduces the CDF to 920\,ms (compared to 1.235\,s for the default option), whereas disabling opportunistic grafting increases the CDF to 1.315\,s.

Clearly, the cold boot attack is a challenging one and requires the whole array of mitigation measures to keep the mesh in a healthy state and meet the propagation latency requirements of the Filecoin and ETH2.0 blockchain networks.

\subsection{Covert Flash Attack}


This attack intends to cover cases where Sybils connect to the network and behave properly for some period of time (2\,min in our experiment) in order to build up score. Then, they execute a coordinated network-wide eclipse attack whereby they stop propagating messages in an attempt to completely disrupt the network. The attack builds on the previous cold boot attack in that Sybils and honest peers connect to the network concurrently. Initial the mesh is dominated by Sybils, who maintain their status by behaving properly until launching the attack.

Again, the whole array of defenses build into GossipSub need to come into play in order to recover from this extremely challenging attack.

Plain GossipSub fails to respond to the attack, which results in message loss and delivery delays of more than 15\,s once the attack is launched. Both Bitcoin's flooding protocol and ETH1.0's \texttt{sqrt(N)}-based pubsub protocol fail to meet the 6\,s deadline, despite the significant redundancy induced, especially in the case of Bitcoin. Bitcoin's protocol results in 3.2M duplicate messages. ETH1.0's \texttt{sqrt(N)} protocol is closer to GossipSub with 230k duplicate messages, against 223k for GossipSub.
The performance of GossipSub is remarkable managing to recover rapidly from the attack and recover the mesh to a healthy state. The maximum delivery latency during recovery tops at about 1\,s, similarly to previous attacks. The p99 of the global propagation latency for GossipSub is only 197\,ms. 

\begin{figure}[h]
\centering
\includegraphics[width=0.3\textwidth]{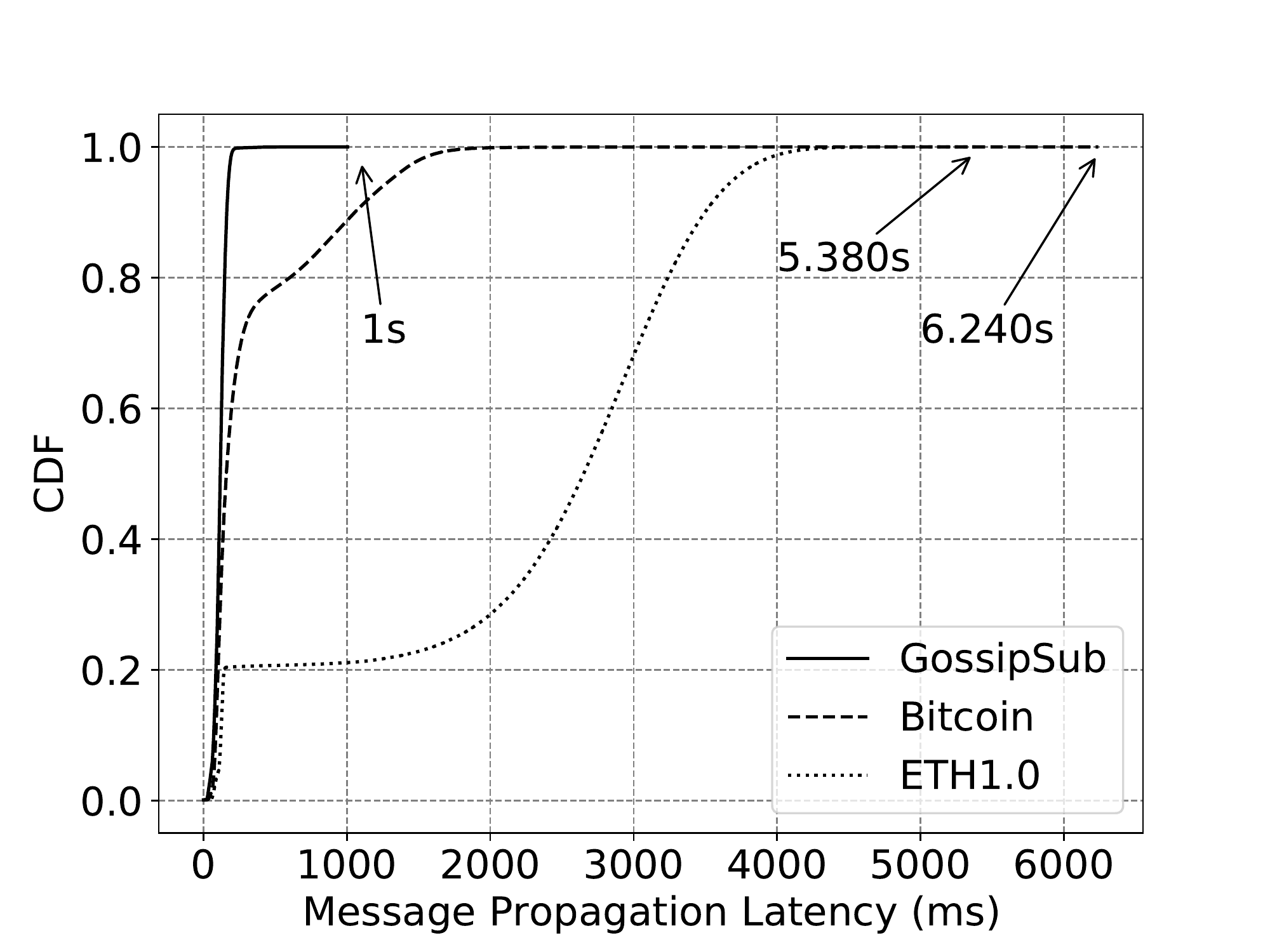}
\caption{Covert Flash Attack}
\label{fig:covert-flash-merged-cdf}
\end{figure}


\subsection{Results Summary}

Having investigated the widest possible range of attacks for message propagation protocols in blockchain environments, we have not identified a case where the GossipSub protocol breaks. Some complex attacks do have an effect on performance, but \textit{do not manage to be successful}. While some parameters of the score function have a clear effect on achieving resiliency (\eg $P_{3a}$ and $P_{3b}$) some others work silently on the background to augment the individual peer view of the mesh. Flood publishing, opportunistic grafting and increased gossip propagation enhance the performance of the protocol significantly under the most adversarial environments. It is this combination of strategies that keep the protocol resilient against all identified cases.

In some attacks, a few of the mitigation strategies are enough to keep the protocol in normal operation. In some of the more complex attacks, the whole arsenal of mitigation strategies has to be utilised in order to keep the effect of the attacks to acceptable levels. We have tested, but are not presenting results that test the resistance of the protocol against collocated Sybils (through the IP Collocation Factor) and SPAM attacks. These are less sophisticated attacks that GossipSub can handle comfortably. 
\section{Concluding Remarks}

We presented GossipSub, a gossip-based, pubsub messaging protocol for message propagation in permissionless blockchain environments. To the best of our knowledge, GossipSub is the first of its kind attack-resistant pubsub protocol that integrates security measures natively at the protocol layer. Responsiveness to dynamic network conditions caused by malicious activity has been the driving philosophy behind the design of GossipSub.
We have not found evidence of a successful attack against GossipSub even in extreme, unrealistic attack scenarios.
\clearpage
\section{Appendix}

\subsection{Bandwidth Requirements \& Cost of Attack}\label{sec:bandwidth-requirements}




In this section, we calculate the bandwidth requirements of GossipSub in order to assess the cost of running Sybil nodes and attack the network.
Throughout our test campaign we have assumed a message rate of 120 msg/sec and message size of 2KBs. While the message size of 2 KB is a reasonable setting, the message rate we have used is very high for both the block and the transactions topic of the Filecoin and the ETH2.0 networks, according to current predictions. This was done on purpose, in order to stress-test the protocol behaviour, but is not an appropriate figure for assessing bandwidth requirements. That said, we will calculate bandwidth requirements for: i) block topic rate of 10 blocks/min, that is, 5 blocks per round with round duration of 30 sec, and ii) transactions topic rate of $\sim$2-4 transactions/second. Where needed we vary these values to provide a holistic picture.

The formula used to calculate bandwidth requirements is:
\\[0.5cm]
\texttt{Bandwidth Req = msg\_rate x msg\_size x connections}
\\[0.5cm]
According to GossipSub's protocol design, a peer will form a mesh with another 8-12 peers (excluding the peers that it is gossiping with). Table~\ref{table-blocks-tx-bandwidth} shows the bandwidth requirement for the Blocks and the Transactions topics.

\begin{center}
\begin{table}[h!]
 \begin{tabular}{||c | c | c ||} 
 \hline
 \multicolumn{3}{|c|}{Data Vol. (GB/month)}\\
 \hline
 Number of & Blocks & Transactions \\
 Connections & (10/min) & (2/sec) \\
  \hline\hline
4 & 3.456 & 41.4 \\ 
\cellcolor{green!25} 8  & \cellcolor{green!25}6.912 & \cellcolor{green!25}82.9 \\
\cellcolor{green!25}12 & \cellcolor{green!25}10.368 & \cellcolor{green!25}124.4 \\
24 & 20.736 & 248.8 \\
48 & 41.472 & 497.6 \\
\cellcolor{red!25}100 & \cellcolor{red!25}86.4 & \cellcolor{red!25}1036 \\
 \hline
\end{tabular}
\caption{GBs/month requirement for Increasing Number of Connections for the Blocks and Transactions Topics, assuming a rate of 10 new messages per minute (\ie 5 per round) for Blocks and 2 messages per second for the Transactions Topic.}
\label{table-blocks-tx-bandwidth}
\end{table}
\end{center}


The green rows in the tables denote bandwidth requirement for GossipSub honest nodes and the red row is the bandwidth requirement for Sybil nodes. 
We make the following observations:
\begin{itemize}[leftmargin=*]
\item Simple calculations reveal that the per-second bandwidth requirement for Sybil nodes (\ie 100 connections each) and for Tx rate of only 2Tx/sec is 3.2Mbps.
\item The bandwidth requirement for honest nodes is high but not prohibitive.
\item Traditionally, botnets are created by compromised user machines, most commonly connected through home connections. Home connection capacities are usually sufficient for normal DDoS attacks and would also be enough for a rate of 2Tx/sec. However, a botnet-based attack would be hard to mount in the FIL/ETH2.0 case in the longer term as the rate of Tx/sec increases. The bandwidth requirement increases linearly with the number of Transactions, which yields a bandwidth requirement of 12.8Mbps for 8 Tx/sec, for instance - prohibitive for the majority of conventional home connectivity compromised nodes.
\item The bottom, red row of Table~\ref{table-blocks-tx-bandwidth} indicates that for 10 blocks/min, 2 Tx/sec and 100 connections per node, a Sybil node requires more than 1 TB of traffic per month. Again, the figure increases linearly with the rate of transactions.
\item According to our AWS-based tests, 1 vCPU with 1GB of memory is enough for a Sybil node to open and operate 100 connections. The bandwidth requirement, however, is non-trivial and hovers in the area between 1 TB and 2 TB.
\end{itemize}

Disregarding the risk of mounting an attack from a contract-based cloud environment, bandwidth-intensive plans come at a non-trivial cost. DigitalOcean offers a 2 TB-compatible plan for \$10/month\footnote{\url{https://www.digitalocean.com/pricing/}}.

Based on this figure, the attacks we tried against our 1000 honest node network, \textit{which ultimately proved unsuccessful, would cost in the order of \$40,000/month}.

We should emphasize that in our test campaign \textit{we have not found an attack that can successfully bend the performance of GossipSub}.
The only somewhat effective attacks were the \textit{cold boot} and \textit{covert flash attacks}, where the attack happens at network formation time. Even in those cases, however, we have proved that the effect on GossipSub's performance was minimal and the network successfully recovered in a short period of time (Section~\ref{sec:cold-boot}).

\subsection{\texttt{D} Value Sensitivity Analysis}\label{sec:sensitivity-analysis}

We carry out a sensitivity analysis for the degree of GossipSub's \texttt{D} value. As discussed earlier, there is a clear tradeoff between the degree of the mesh and the performance of the protocol in terms of bandwidth requirements and message propagation latency (see Figure~\ref{fig:tradeoffs}). In Table~\ref{table-d-values} we present a subset of the range of \texttt{D} values that we experimented with.
\begin{center}
\begin{table}[H]
\begin{tabular}{|p{2.5cm}||P{0.9cm}|P{1cm}|P{1cm}|P{1cm}|}
 \hline
 \multicolumn{5}{|c|}{GossipSub \texttt{D}-Value Settings} \\
 \hline
 & \texttt{Lo\_D} & \cellcolor{red!50} \texttt{def\_D} & \texttt{Hi\_D} & \texttt{VHi\_D} \\
 \hline
 \rowcolor{gray!50}\texttt{D}: Degree of Mesh & 4 & 8 & 16 & 32 \\
 \texttt{D\_low} (\texttt{D} -25\%) & 3 & 6 & 12 & 24 \\
 \texttt{D\_score} (\texttt{D} -25\%) & 3 & 6 & 12 & 24 \\
 \texttt{D\_high} (\texttt{D} +50\%) & 6 & 12  & 24 & 48 \\
\hline
\end{tabular}
\caption{Range of \texttt{D} Values Tested}
\label{table-d-values}
\end{table}
\end{center}
Lower degree for the mesh (\texttt{Lo\_D=4}) results in lower number of duplicate messages ($\sim$100k) (approximately half of those for the default \texttt{D} values, $\sim$227k), but also higher message propagation latency - 245ms for ``Low \texttt{D}'', compared to 185ms for the ``Default \texttt{D}'' value (Table~\ref{table-d-sensitivity}). Increasing the degree of the mesh to \texttt{Hi\_D=16} decreases message propagation latency (by $\sim$25ms) but results in more than 100\% increase in the number of duplicate messages - $\sim$480k in case of ``High \texttt{D}'', compared to $\sim$227k for the ``Default \texttt{D}''. Increasing the \texttt{D} value further to 32 makes virtually no difference in terms of message propagation latency, but increases the number of duplicate messages by 100\%, as shown in Table~\ref{table-d-sensitivity}.
\begin{center}
\begin{table}[H]
\begin{tabular}{|p{2cm}||P{1cm}|P{1.1cm}|P{1cm}|P{1.5cm}|}
 \hline
 \multicolumn{5}{|c|}{GossipSub \texttt{D}-Value Sensitivity Analysis - Baseline Test} \\
 \hline
 & Low \texttt{D} (D=4) & Default \texttt{D} (D=8) & High \texttt{D} (D=16) & Very High \texttt{D} (D=32) \\
 \hline
 Max Latency & 245ms & 185ms & 161ms & 160ms \\
 p99 PDF & 175ms & 139ms & 117ms & 113ms \\
 Duplicates* & 100k & 227k & 480k & 995k \\
\hline
\end{tabular}
\caption{Sensitivity Analysis Results}
\label{table-d-sensitivity}
\end{table}
\end{center}
The number of duplicate messages (and as a consequence the volume of redundant traffic) increases exponentially with the degree of the mesh. This is also the case with the amount of traffic that the protocol produces as a result of higher-degree connectivity. In Table~\ref{table-increasing-tx-bandwidth}, we present the bandwidth requirements for increasing number of transactions per second for GossipSub (degree $\sim$8), ETH1.0's \texttt{sqrt(N)}-based protocol (degree $\sim$32 for 1000 nodes) and Bitcoin's broadcast protocol (degree $\sim$133 for public Bitcoin nodes).

\begin{center}
\begin{table}[h!]
 \begin{tabular}{||c | c | c |c ||} 
 \hline
 \multicolumn{4}{|c|}{Data Vol. (GB/month)}\\
 \hline
 Tx/sec & GossipSub & ETH1.0 & Bitcoin \\ [0.5ex]
 \hline\hline
2 & 124.4 & 331.7 & 1378\\ 
4  & 248.8 & 663.5 & 2757 \\
8 & 497.6 & 1327 & 5515 \\
16 & 995.3 & 2654 & 11031 \\
32 & 1990 & 5308 & 22063 \\
64 & 3981 & 10616 & 44126 \\
 \hline
\end{tabular}
\caption{GB/month for an increasing number of Transactions for GossipSub (\texttt{D = 12} connections), Ethereum (32 connections for a network of 1k nodes), Bitcoin (133 connections in total, 125 inbound, 8 outbound)}
\label{table-increasing-tx-bandwidth}
\end{table}
\end{center}

Based on these observations, we deemed the value of \texttt{D=8} as a reasonable setting and have used this by default. Although different settings might be more suitable for different environments, this setting has proved to be ideal for the wide range of scenarios tested.


\subsection{Testground Description}\label{sec:testground-description}

The Testground runtime builds on a traditional daemon/client architecture, communicating over HTTP. This architecture is flexible enough to run the daemon within a shared cluster, with many users, developers, and integrations (\eg GitHub Actions) hitting the same daemon to schedule build and run jobs.

Testground includes a number of attractive features for experimentation and reproducibility of results.
\begin{itemize}
    \item \textit{Distributed coordination API.} Redis-backed lightweight API offering synchronisation primitives to coordinate and orchestrate distributed test workloads across a fleet of nodes.
    \item \textit{Tunable Network Conditions.} Bandwidth, latency, jitter and packet-loss can be tuned to emulate realistic network conditions.
    \item \textit{AWS cluster setup scripts.} Testground comes with scripts to setup AWS clusters and reproduce results from previous tests.
    \item \textit{Composition-based declarative jobs.} Tailored test run scripts to compose scenarios in a declarative manner.
    \item Results and diagnostics. When a test run completes debugging logs and results are collected for easy post-processing using Python Jupyter notebooks.
\end{itemize}

\clearpage 
\bibliographystyle{acm}
\bibliography{gossipsub-bib} 

\end{document}